\documentclass{article}
\usepackage{spconf,amsmath,graphicx,hyperref}

\usepackage{cite}
\usepackage{multirow}
\usepackage{tabularx}
\usepackage{booktabs}  
\usepackage{array}

\usepackage{cleveref}

\def \etal {et al.~}

\title{Finding My Voice: Generative Reconstruction of Disordered Speech for Automated Clinical Evaluation}

\name{Karen Rosero$^{1}$, Eunjung Yeo$^{1,2}$, David R. Mortensen$^{1}$, Cortney Van’t Slot$^{3}$, Rami R. Hallac$^{3,4}$, Carlos Busso$^{1}$}
\address{$^{1}$ Language Technologies Institute, Carnegie Mellon University, PA, USA;\\ $^{2}$ Department of Computer Science, University of Texas at Austin, TX, USA; \\ $^{3}$ Department of Plastic Surgery, University of Texas Southwestern Medical Center, TX, USA; \\ $^{4}$ Analytical Imaging and Modeling Center, Children's Health, TX, USA\\\{kroseroj, eyeo2, dmortens, busso\}@andrew.cmu.edu, \{cortney.vantslot, rami.hallac\}@childrens.com}

\begin{document}
\ninept 
\maketitle
\begin{abstract}
We present ChiReSSD, a speech reconstruction framework that preserves children speaker's identity while suppressing mispronunciations. Unlike prior approaches trained on healthy adult speech, ChiReSSD adapts to the voices of children with \textit{speech sound disorders} (SSD), with particular emphasis on pitch and prosody. We evaluate our method on the STAR dataset and report substantial improvements in lexical accuracy and speaker identity preservation. Furthermore, we automatically predict the phonetic content in the original and reconstructed pairs, where the proportion of corrected consonants is comparable to the \textit{percentage of correct consonants} (PCC), a clinical speech assessment metric. Our experiments show Pearson correlation of $\rho$ = 0.63 between automatic and human expert annotations, highlighting the potential to reduce the manual transcription burden. In addition, experiments on the TORGO dataset demonstrate effective generalization for reconstructing adult dysarthric speech. Our results indicate that disentangled, style-based TTS reconstruction can provide identity-preserving speech across diverse clinical populations.

\end{abstract}

\section{Introduction}
\label{sec:intro}

Speech disorders create substantial barriers to daily communication, with disordered speech exhibiting deviations across acoustic, phonetic, and prosodic dimensions that reduce intelligibility and quality of life \cite{bruneel2019relationship, enderby2013disorders}. Disordered speech reconstruction, the task of generating more intelligible utterances while preserving speaker identity, has emerged as a promising option to mitigate these challenges. For example, reconstructed speech can enable the use of general-purpose \textit{automatic speech recognizers} (ASRs) \cite{wang2020end, yang20, prananta2022effectiveness}, support personalized \textit{voice communication aids} (VOCAs) \cite{aihara2017phoneme}, and \textit{pronunciation training} \cite{yang20}, where preserving speaker’s identity is important \cite{hosaka2021neural, Powers2018OwnVoiceProcessing}.

Despite this promise, previous studies on disordered speech reconstruction face two limitations. First, most models are trained on healthy speech, leading to degraded performance on disordered speech \cite{el2025unsupervised}. Second, existing approaches largely target adults, leaving child speech underexplored \cite{zhang24d_interspeech}. For this reason, reconstruction for children with speech disorders is particularly challenging, because their articulation and prosodic patterns differ greatly from those of healthy adults. Addressing this gap is important because effective reconstruction can support children's communication, helping to reduce the long-term social and academic difficulties \cite{hitchcock2015social}. 

We propose ChiReSSD, a StyleTTS2-based framework \cite{Li_2024_2} that disentangles acoustic and prosodic style embeddings. Our contributions are: (i) a novel adaptation method of StyleTTS2 for disordered speech reconstruction, which suppress pronunciation-linked acoustic style while preserving speaker identity, (ii) a comprehensive evaluation combining objective metrics (speaker similarity, lexical accuracy) with clinical intelligibility (PCC),  (iii) evidence that reconstruction improves PCC and that an automatic consonant accuracy metric correlates with clinicians ($\rho$ = 0.63), and (iv) demonstration that ChiReSSD generalizes to adult dysarthric speech, indicating broader clinical applicability. 

\section{Related Works}\label{sec:related_works}

Early assistive communication technologies for individuals with speech disorders relied on voice banking and VOCAs, which synthesized personalized voices from premorbid recordings \cite{aihara2017phoneme}. While effective for identity preservation, these systems require data collection prior to the onset of impairment, thereby limiting their feasibility for individuals seeking support after disorder onset.

More recently, \emph{Voice conversion} (VC) has become the predominant approach for disordered speech reconstruction. GAN-based VC improved intelligibility but offered limited style control \cite{chen19b_interspeech, yang20, purohit2020intelligibility, prananta2022effectiveness, chu2023dgan}, while neural encoder-decoder methods attempted to disentangle content, prosody, and speaker identity \cite{biadsy2019parrotron, wang2020end, doshi2021extending, chen2021conformer}. However, these approaches often rely on paired corpora, which are scarce in clinical contexts. Although zero-shot VC has been explored to relax parallel data requirements \cite{liu2024two}, it remains difficult to improve intelligibility without also transferring residual pathological traits.

Leveraging pre-trained TTS has emerged as a complementary pathway, synthesizing speech from text while conditioning on speaker style. Recent adaptations of Parler-TTS and XTTS-v2 for dysarthric reconstruction \cite{sanchez25_interspeech, szekely25_interspeech} demonstrate data efficiency of these approaches. However, most existing methods also adapt holistically to individual speakers, which risks reproducing disordered patterns. Furthermore, existing work has focused primarily on adult dysarthria, leaving pediatric speech disorders largely unexplored despite their distinct characteristics.

\vspace{-0.2cm}
\section{Methodology}
\label{sec:methods}
\vspace{-0.1cm}


We propose the \emph{Children’s Reconstructed Speech for SSDs} ChiReSSD method, a framework that adapts StyleTTS2 \cite{Li_2024_2} for the reconstruction of disordered child speech. ChiReSSD leverages StyleTTS2’s disentangled acoustic and prosodic style embeddings to selectively reduce the influence of pathological acoustic patterns while retaining prosody and speaker identity. This adaptation is motivated by the observation of Rosero \etal \cite{Rosero_2025_interspeech} that acoustic embeddings may encode mispronunciations as part of style, leading to error reproduction in style-based TTS.

\begin{figure}[t]
    \centering
\includegraphics[width=\columnwidth]{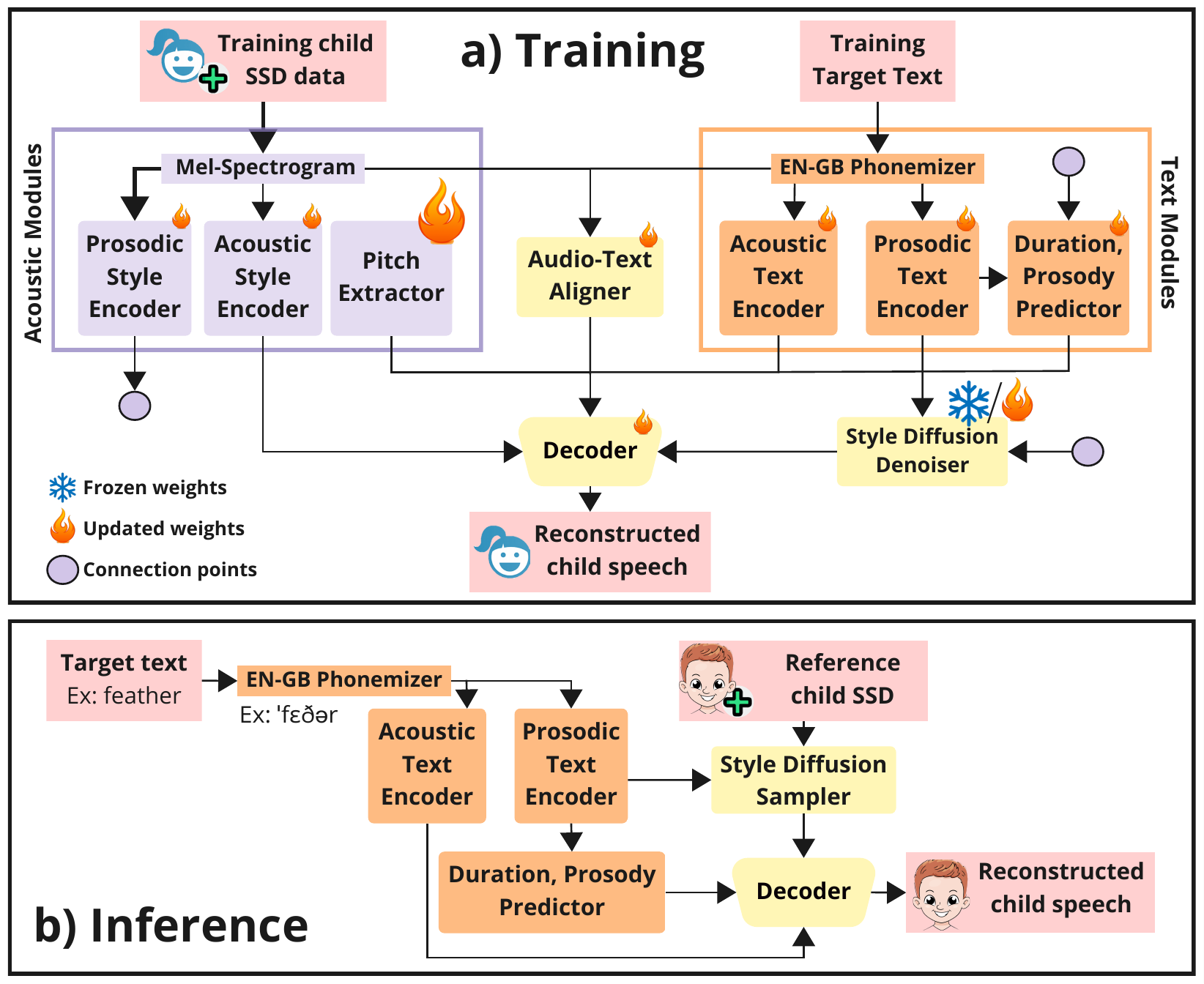}
    \vspace{-0.5cm}
    \caption{ChiReSSD architecture builds upon StyleTTS2. Modules fine-tuned with different loss weights are marked with flames. Partial fine-tuning in style diffusion is denoted with the frozen and flame symbols.}
    \label{img:chiressd}

\end{figure}

As illustrated in Figure~\ref{img:chiressd}a, acoustic modules (purple box) are trained on child SSD data. The framework fine-tunes acoustic and prosodic style encoders as well as a pitch extractor. The latter is of particular interest since StyleTTS2 was originally trained on adult speech and must be adapted to better capture the higher pitch ranges and prosodic patterns of child speech. To this end, we increased the weight of the pitch reconstruction loss and partially froze the diffusion model parameters.

For the text modules (orange box), target transcriptions are converted into IPA sequences using a British English phonemizer \footnote{Specifically, we use the `en-gb' phonemizer available at https://github.com/bootphon/phonemizer based on empirical grounds.}, providing phonemic representations aligned with the accent of our target population (Section \ref{ssec:datasets}). 
These phonemes are then used to train the acoustic and prosodic text encoders. 
The acoustic and text modules interact through an audio-text aligner, and contribute to the estimation of prosody and speech duration. Both modalities also condition the style diffusion denoiser during training. 
Finally, the acoustic and text modules, together with alignment and style representations, guide the HiFi-GAN decoder to generate reconstructed child speech. These updated weights now represent our trained ChiReSSD model.


The inference process is illustrated in Figure~\ref{img:chiressd}b. It begins with the extraction of style embeddings from a reference sample of an unseen child with SSD. These embeddings are computed on a per-subject basis and can subsequently be used to generate reconstructed speech for any target text. Importantly, this design enables one-shot style transfer without requiring parallel data: a reference recording of only four seconds is sufficient, even if it does not contain the target utterance. The target lexical content is encoded at the phoneme level and passed through the acoustic and prosodic text encoders, which estimate duration and prosody. 

During generation, we control two guidance parameters: $\alpha$, which scales the contribution of the acoustic text encoder, and $\beta$, which regulates prosodic similarity to the source style. Finally, the decoder generated the reconstructed speech with guidance of the  style diffusion sampler, the acoustic and prosodic text encoders and the duration and prosody predictors. \\


\noindent
\textbf{Automated Clinical Evaluation}: Generating speaker-preserving reconstructed speech signals opens new research directions to identify localized speech production errors. This study explores the use of ChiReSSD for the automatic prediction of \textit{percentage of correct consonants} (PCC). Starting from each original SSD sample and its reconstructed counterpart, we examine the phonetic differences between the two. To obtain automatic IPA transcriptions, we employ \emph{wav2vec2-lv-60-espeak-cv-ft}, a universal phone recognizer that processes input speech. From these predictions, we extract only consonants to align with established clinical speech evaluation metrics. This focus is motivated by the high prevalence of consonant production difficulties in the pediatric population. Finally, we compute the Levenshtein distance between the consonant sequences of the original and reconstructed samples and assess the correlation between this automatically derived measure and expert clinical annotations.

\section{Experimental Setup}

All components in ChiReSSD are initialized from pre-trained StyleTTS2 checkpoints. ChiReSSD was fine-tuned for four epochs in total to avoid overfitting to SSD-specific pronunciations, with the diffusion model parameters frozen for the first two epochs and an increased weight of the pitch reconstruction loss. Model training required an A100 GPU with 80GB of VRAM for a batch size of 4. 

During generation, we set $\alpha=0.8$ to enforce less similarity to the mispronunciation patterns and $\beta=0.6$ to balance prosody preservation with natural variation. Generation is performed with 10 diffusion steps and an embedding scale of 1, achieving a compromise between computational efficiency and fidelity to both style and content. 

\vspace{-0.3cm}
\subsection{Datasets}\label{ssec:datasets}

We focus on children with SSD using the UltraSuite and STAR corpora, both involving speakers of English with a Central Scottish accent. To assess the generalization of our approach, we also evaluate on TORGO, which contains recordings of adult speakers with dysarthria.

\noindent
\textbf{UltraSuite Corpus}: UltraSuite \cite{Eshky_2018} contains audio recordings from pediatric speech therapy sessions with children aged 5–12 having various speech disorders, such as phonological delay/disorder, inconsistent realizations, articulation difficulties, and childhood apraxia of speech. We use data from subjects with ground-truth transcriptions: UXSSD (subjects 1,4), UPX (subjects 1–20), and UX2020 (subjects 1–37). We manually clean the data to remove therapist segments and correct alignments. We refer to this curated subset as the Children-SSD (CSSD) dataset, which contains 2,429 utterances that will be used exclusively for training.
 
\noindent
\textbf{Speech Therapy Animation and Imaging Resource (STAR)}: 
For evaluation, we use the STAR database of disordered speech errors \cite{lawson2023star_1, lawson2023star_2}, built from six pre-existing datasets collected for clinical and non-clinical purposes. STAR provides phonemic transcriptions for a Central Scottish rhotic accent (target forms) and clinician-verified phonetic transcriptions reflecting childrens' actual mispronunciations. The dataset comprises 234 utterances from six children, designed to elicit specific sounds that enable detailed comparison between intended and produced phonemes.  

\noindent
\textbf{TORGO Dataset}: To evaluate cross-disorder generalization, we use TORGO \cite{rudzicz2012torgo}, a dataset of dysarthric and control adult speakers of Canadian English. We use 1,123 samples of dysarthric speech, which are categorized per severity for mild (F03, F04, M03), moderate (F01, M05), and severe (M01, M02, M04) levels.

\subsection{Baselines}

We refer to our first baseline as \textit{Standard TTS}, which is obtained with StyleTTS2 trained with the LJSpeech single-speaker dataset. The lexical content of samples in the STAR dataset guides the generation. This baseline represents a TTS approach without style transfer or personalization for children speech. Our second baseline is referred to as \textit{One-shot TTS}, for which the original child SSD utterance of the STAR dataset is used to extract the style representation in a one-shot setting of StyleTTS2. This approach is expected to produce speech as similar as possible to the reference utterance without having been specifically trained on children speech. 

\subsection{Evaluation metrics}\label{sec:metrics}

\noindent
\textbf{Objective evaluation}: We consider metrics for the preservation of speaker identity, computed between the original and reconstructed samples. \textit{Speaker similarity} is computed between 256-dimensional embeddings extracted using Resemblyzer \cite{wan2018generalized}. To assess prosodic fidelity, we extract the \textit{fundamental frequency} $F_0$ (perceived pitch) with the YIN algorithm \cite{de2002yin}. We also report the semitone difference, expressed as \mbox{$12 \cdot \log_2(F_{0\text{ rec}} / F_{0\text{ ref}})$}, and the relative pitch deviation in percentage. 

To assess the \textit{lexical and phonetic accuracy} of the reconstructed speech, we use the \texttt{whisperX large‑v2} model \cite{bain23_interspeech} for transcription. Then, the \textit{word error rate} (WER) and \textit{character error rate} (CER) are calculated based on the ground-truth transcripts. In addition, we evaluate the model confidence by supplying the ground-truth word sequence and measuring how confident \texttt{whisperX large‑v2} is that each recognized word matches the expected transcript. \\

\noindent
\textbf{Clinical evaluation}: To assess speech accuracy from a clinical perspective, we adopt the PCC, a well-established metric for quantifying the severity of speech sound disorders \cite{shriberg1994developmental}. A certified \textit{speech-language therapist} (SLT) specializing in childhood SSDs annotated 21 pairs of original and reconstructed samples, providing the number of correctly produced consonants per utterance. Each sample was evaluated independently to avoid bias.

\section{Experiments and Results}
\label{sec:experiments} 

We evaluate our approach across complementary perspectives. Section~\ref{sec:skr_sim} analyzes speaker similarity to assess preservation of identity, while Section~\ref{sec:wer_cer} reports lexical and phonetic accuracy. In Section~\ref{sec:clinical_eval}, we compare our computational metrics with clinical evaluations. Finally, Section~\ref{sec:torgo} investigates the generalization of our method to adult dysarthric speech. 

\subsection{Preservation of Speaker Identity} \label{sec:skr_sim}

Preserving the speaker’s identity in reconstructed speech is central to our framework. Table~\ref{tab:SPKID} reports the similarity between embeddings of original STAR samples and their reconstructions. Our ChiReSSD approach achieves a similarity score of 0.62, surpassing both One-shot TTS and Standard TTS. A threshold of 0.6 has been reported as acceptable for speaker verification tasks \cite{zhang_2022_sig}, suggesting that ChiReSSD retains sufficient speaker-specific traits for reliable identification. In contrast, Standard TTS yields only 0.40, reflecting the absence of style transfer despite sharing lexical content with the originals. One-shot TTS benefits from reference embeddings but underperforms due to its lack of training on child voices, often producing reconstructions with adult female pitch contours. These results highlight the necessity of our child-specific adaptation.

To disentangle lexical overlap from acoustic resemblance, we further analyze perceived pitch similarity. ChiReSSD achieves the lowest $F_0$ difference (19.98\%, corresponding to 3.83 semitones), indicating close alignment with original pitch distributions. Prior psychoacoustic and speech-perception studies suggest that pitch differences up to about 3 semitones are often imperceptible or minimally disruptive in voice identity perception \cite{flaherty2021independent}. In contrast, One-shot TTS exhibits larger deviations, consistent with its tendency to approximate child voices using adult pitch ranges. Standard TTS shows the greatest discrepancy (39.95\%, ~6.09 semitones), well above typical perceptual thresholds, which likely contributes to a less convincing preservation of speaker identity. Together, these findings show that ChiReSSD not only stays within perceptually acceptable semitone bounds, but outperforms the other approaches in maintaining both identity and child-matched prosody.

\begin{table}[t]
 \caption{Preservation of speaker identity metrics. Mean values computed for samples of the STAR dataset are presented. }
\label{tab:SPKID}
\begin{tabular}{cccccc} \toprule \textbf{Method} & \textbf{Similarity $\uparrow$} & \textbf{$F_0$ diff. $\downarrow$} & \textbf{Semitone diff. $\downarrow$} \\ \toprule
ChiReSSD           &  \textbf{0.62*} &  \textbf{19.98\%*}     &     \textbf{3.83*}                             \\
One-shot TTS &  0.52 &  22.58\%    &      4.14                           \\
Standard TTS            &  0.40  &  39.95\%     &      6.09                   \\ 

\bottomrule
\end{tabular}\\
\footnotesize{(*) Denotes a statistically significant improvement based on a t-test compared to the baselines ($p < 0.05$).}
\end{table}

\subsection{Lexical and Phonetic Accuracy of Reconstructed Speech} \label{sec:wer_cer}

ASR on children’s speech is notoriously difficult due to immature articulation patterns, and becomes even more challenging in the presence of SSDs. Table \ref{tab:wer_cer} reports WER and CER obtained with \texttt{whisperX large-v2}. As expected, original STAR samples without reconstruction yielded very high error rates (WER 0.84, CER 0.56), reflecting the acoustic and phonetic deviations from canonical pronunciations. 

Reconstruction substantially reduced error rates. Our proposed ChiReSSD achieved a 48\% relative reduction in CER (0.56 $\rightarrow$ 0.29) and a 42\% relative reduction in WER (0.84 $\rightarrow$ 0.49). One-shot TTS showed similar reductions, while Standard TTS (benefiting from adult-like synthesis) unsurprisingly yielded the lowest WER and CER (0.34 and 0.11, respectively). These results indicate that ChiReSSD reduces phonetic distortions. 
We expect that residual errors are partly explained by \texttt{whisperX}’s limitations on non-sentential word sequences present in STAR. 

To test reconstruction in a more naturalistic setting, we synthesized 234 sentences from the LibriTTS test set using STAR-derived style embeddings. In this case, ChiReSSD reached CER and WER values of 0.06 and 0.15, respectively, aligning with state-of-the-art reports on child ASR \cite{fan24b_interspeech}. This result confirms that the  error rates on STAR are largely attributable to atypical non-sentential word sequences and not to limitations of our reconstruction pipeline.  

Confidence scores from \texttt{whisperX} further corroborate these findings. STAR original samples yielded extremely low confidence (0.28), while reconstruction improved scores across all methods. ChiReSSD achieved 0.71, slightly lower than One-shot TTS (0.75) and Standard TTS (0.85). This gap reflects the fact that ChiReSSD produces child-like speech patterns that pose greater challenges to ASR models trained primarily on adult speech. Nevertheless, when applied to LibriTTS material, ChiReSSD reached a confidence of 0.83, confirming its robustness in reconstructing natural child speech while preserving stylistic characteristics.

\begin{table}[t]
\caption{CER, WER, and confidence computed for the indicated datasets using different methods considered in this study.}
\label{tab:wer_cer}
\begin{tabular}{ccccc}
\toprule
\textbf{Method}   &   \textbf{Dataset}       & \textbf{CER } & \textbf{WER} & \textbf{Confidence} \\ \toprule
No reconstruction   & STAR          & 0.56               & 0.84    &   0.28 \\
One-shot TTS* &  STAR  & 0.30               & 0.52    &   \textbf{0.75} \\
ChiReSSD* &  STAR          & \textbf{0.29}               & \textbf{0.49}    &   0.71 \\
Standard TTS* &  STAR           & 0.11               & 0.34    &   0.85 \\
ChiReSSD* &  LibriTTS           & 0.06               & 0.15    &   0.83 \\ \bottomrule
\end{tabular}\\
\footnotesize{(*) Denotes a statistically significant improvement based on a t-test compared to the original samples of STAR without reconstruction ($p < 0.05$).}
\end{table}

\subsection{Correlation with Clinical Evaluations}\label{sec:clinical_eval}
For a clinically grounded assessment of speech accuracy, we obtained PCC from prediction of IPAs and clinical evaluations through the process described in Section \ref{sec:methods}. 
We obtained a PCC of 87.44\% for the reconstructed samples while for the original samples, a PCC was 68.14\%, indicating that ChiReSSD substantially improves consonant accuracy in disordered child speech.

In parallel, we processed the manual annotations obtained from an SLT for 21 pairs of original and reconstructed samples, which include the number of correct consonants and the total number of consonants in the input. 
According to the clinical evaluation, the PCC in the original samples was 70.13\%, and improved to 94.81\% after ChiReSSD reconstruction. In contrast, the computational analysis presented a PCC of 50.65\% for the same group of original samples and improved to 92.21\% with the ChiReSSD reconstruction. Despite not perfect, we observe certain alignment between the computational and medical evaluations.

Finally, we computed the Pearson correlation coefficient between the automatically computed consonant differences and those derived from human annotations. We observed a Pearson correlation of $\rho$ = 0.63 between the automatic and manual annotations, indicating a moderate yet meaningful alignment. This result highlights a significant opportunity to reduce the effort involved in manual phonetic transcription for PCC calculation. Specifically, this goal can be achieved by automatically comparing the number of corrected characters in the reconstructed speech samples with the original IPA characters predicted by the system.

\subsection{Generalization to Dysarthric Adult Speech} \label{sec:torgo}

To evaluate the generalization capability of ChiReSSD beyond childhood SSDs, we conducted experiments on the TORGO dataset, which contains adult speakers with dysarthria. We used a general English phonemizer for the transcriptions. Table~\ref{tab:torgo} reports the same metrics used for children speech. The results demonstrate that our approach substantially improves lexical accuracy across severity levels: CER is consistently reduced to below 0.03, even for severe dysarthria (from 0.40 to 0.02), while WER decreases from 0.61 to 0.06 in the severe group. Confidence scores also increase across all severity groups, approaching 0.80 for severe dysarthria.  

Notably, speaker identity preservation remains robust, with similarity values of 0.74–-0.77, far above the 0.6 threshold commonly cited for speaker identification. This result indicates that ChiReSSD captures prosodic and stylistic cues while avoiding the reproduction of pathological articulation patterns. Pitch difference analysis further reveals a marked reduction compared to Standard TTS, particularly for severe cases (31.39\% vs. 39.53\%), suggesting more faithful reproduction of the target prosody. These findings highlight the versatility of our approach to generalize effectively to adult dysarthric speech, another population with distinct acoustic deviations. This cross-disorder adaptability underscores the potential of disentangled, style-based TTS frameworks to provide personalized, intelligible, and identity-preserving speech reconstruction for diverse clinical populations.  

\begin{table}[t]
\caption{Metrics computed for subjects with dysarthria of the TORGO dataset using the ChiReSSD and Standard TTS.}
\label{tab:torgo} 
\centering
\renewcommand{\arraystretch}{0.7}
\small
\begin{tabularx}{\columnwidth}{
  >{\centering\arraybackslash}p{1.2cm}  
  >{\centering\arraybackslash}p{1.3cm} 
  >{\centering\arraybackslash}p{1.1cm}   
  >{\centering\arraybackslash}p{1.4cm}
  >{\centering\arraybackslash}X }
\toprule
\textbf{Metric} & \textbf{Severity} & \textbf{Original} & \textbf{ChiReSSD} & \textbf{Stand TTS } \\
\toprule
\multirow{3}{*}{CER} 
 & Mild      & 0.06 & 0.01 & 0.01 \\
 & Moderate  & 0.23 & 0.02 & 0.02 \\
 & Severe    & 0.40 & 0.02 & 0.02 \\
\midrule
\multirow{3}{*}{WER} 
 & Mild      & 0.13 & 0.05 & 0.04 \\
 & Moderate  & 0.37 & 0.08 & 0.07 \\
 & Severe    & 0.61 & 0.06 & 0.05 \\
\midrule
\multirow{3}{*}{Confidence} 
 & Mild      & 0.79 & 0.84 & 0.84 \\
 & Moderate  & 0.72 & 0.82 & 0.84 \\
 & Severe    & 0.70 & 0.80 & 0.83 \\
\midrule
\multirow{3}{*}{\shortstack{Speaker \\ similarity}} 
 & Mild      & ---    & 0.77 & 0.55 \\
 & Moderate  & ---    & 0.76 & 0.53 \\
 & Severe    & ---    & 0.74 & 0.43 \\
\midrule
\multirow{3}{*}{\shortstack{$F_0$ \\ difference}} 
 & Mild      & ---    & 15.60\% & 23.45\% \\
 & Moderate  & ---    & 21.76\% & 25.07\% \\
 & Severe    & ---    & 31.39\% & 39.53\% \\
\bottomrule
\end{tabularx}\\
\footnotesize{``-'' indicates original condition (self-comparison omitted).}
\end{table}

\section{Conclusions}
\label{sec:conclusions}

We have presented ChiReSSD, a style-based TTS reconstruction framework tailored to children with SSD, which emphasizes preserving speaker identity while correcting mispronunciations. Our extensive evaluation shows that ChiReSSD substantially reduces lexical and phonetic distortion, achieving large relative gains in CER and WER over baseline reconstruction approaches. It also preserves speaker identity above commonly cited thresholds for speaker verification tasks, supported by both embedding similarity and pitch analyses. 

Clinical evaluation confirmed the relevance of these gains: automatic estimations of PCC for the original and reconstructed speech correlated in 0.63 points with expert annotations, pointing to the feasibility of automating clinical metrics aided by our reconstruction approach. Moreover, results on the TORGO dataset demonstrated that ChiReSSD generalizes to adult dysarthria, consistently improving WER and CER and keeping prosodic fidelity across severity levels.  

In future work, we aim to further reduce residual phonetic errors by integrating more robust phoneme-aware loss functions. We also plan to expand the framework to end-to-end training that includes ASR directly from the original SSD speech.

\bibliographystyle{IEEEbib}
\bibliography{strings, refs}

\end{document}